\documentclass[journal]{IEEEtran}

\usepackage{amsmath,amssymb,amsfonts}
\usepackage{bm}
\usepackage[english]{babel}
\usepackage{amsthm}
\usepackage{multirow}
\usepackage{tabularx}
\usepackage{color}
\usepackage{cite}
\usepackage{silence}
\usepackage{caption}
\usepackage{blindtext}
\usepackage{pstool}
\usepackage{psfrag}
\usepackage{lettrine}
\usepackage{float}
\usepackage{makecell}
\usepackage[export]{adjustbox}
\usepackage{graphicx}
\usepackage{booktabs}
\usepackage[dvips]{epsfig}
\usepackage{boxedminipage}
\usepackage{stfloats}
\usepackage{lipsum}
\usepackage[ruled, linesnumbered]{algorithm2e}
\usepackage{xcolor}
\usepackage{braket}

\SetKwInOut{Input}{Input}
\SetKwInOut{Output}{Output}

\usepackage{scalerel}
\usepackage{tikz}

\usepackage[bookmarks=false]{hyperref}

\DeclareCaptionLabelSeparator{periodspace}{.\quad}
\IEEEoverridecommandlockouts

\begin{document}

\title{Quantum-MUSIC: Multiple Signal Classification for Quantum Wireless Sensing
\thanks{
This work was supported by the National Research Foundation of Korea (NRF) grant funded by the Korea government (MSIT) (No. RS-2024-00409492).
\newline \indent Hanvit Kim, Hyunwoo Park, and Sunwoo Kim are with the Department of Electronics and Computer Engineering, Hanyang University, Seoul, 04763, South Korea (email: dante0813@hanyang.ac.kr; stark95@hanyang.ac.kr; remero@hanyang.ac.kr).
}}

\author{
\IEEEauthorblockN{Hanvit Kim,~\IEEEmembership{Graduate Student Member,~IEEE}, Hyunwoo Park,~\IEEEmembership{Graduate Student Member,~IEEE}, and Sunwoo Kim,~\IEEEmembership{Senior Member,~IEEE}}}

\maketitle

\begin{abstract}
This paper proposes a Quantum-MUSIC, the first multiple signal classification (MUSIC) algorithm for quantum wireless sensing of multi-user. Since an atomic receiver for quantum wireless sensing can only measure the magnitude of a received signal, sensing performance degradation of traditional antenna-based signal processing algorithms is inevitable. To overcome this limitation, the proposed algorithm recovers the channel information and incorporates the traditional MUSIC algorithm, enabling the sensing of multi-user with magnitude-only measurement. Simulation results showed that the proposed algorithm outperforms the existing MUSIC algorithm, validating the superior potential of quantum wireless sensing.
\end{abstract}

\begin{IEEEkeywords}
Quantum wireless sensing, atomic receiver, signal processing
\end{IEEEkeywords}

\author{
\IEEEauthorblockN{Hanvit Kim,~\IEEEmembership{Graduate Student Member,~IEEE}, Hyunwoo Park,~\IEEEmembership{Graduate Student Member,~IEEE}, and Sunwoo Kim,~\IEEEmembership{Senior Member,~IEEE}}}

\maketitle

\section{Introduction}
Quantum wireless system is envisioned to bring a new paradigm shift to wireless communication and sensing\cite{QuantumWirel}. Unlike the traditional wireless system that exploits radio frequency (RF) components such as amplifiers and mixers, the quantum wireless system exploits an atomic receiver for wireless signal reception and processing \cite{atomic_receiver}. As shown in Fig. 1, the atomic receiver leverages the \textit{Rydberg atoms}, the high-level energy atoms with the principal quantum number higher than 20, for detecting the RF signal's amplitude, frequency, and phase \cite{Quantum_Phase}. The extreme sensitivity of Rydberg atoms to external electromagnetic (EM) fields is expected to bring several noticeable advantages, such as high sensing granularity, robustness against thermal noise, and ultra-wide spectrum sensing\cite{Rydberg_Atoms1}.

To fully exploit the aforementioned potential, theoretical modeling and developing signal processing algorithms for atomic receivers are promising research directions \cite{Rydberg_Receiver, Atomic-MIMO, IQ}. In \cite{Atomic-MIMO}, the atomic receivers are incorporated into the multiple-input-multiple-output (MIMO) for the signal detection of multi-user. Here, a biased Gerchberg-Saxton (GS) algorithm was proposed to solve the non-linear biased phase retrieval (PR) problem induced by the magnitude-only measurement of the atomic receiver. Authors in \cite{IQ} proposed a MIMO precoding design for the atomic receiver, where independent signal processing of I/Q signal data is exploited, and the information-theoretical limit of the atomic MIMO system.

\begin{figure}[t]
    \begin{center}
        \includegraphics[width=0.95\columnwidth,center]{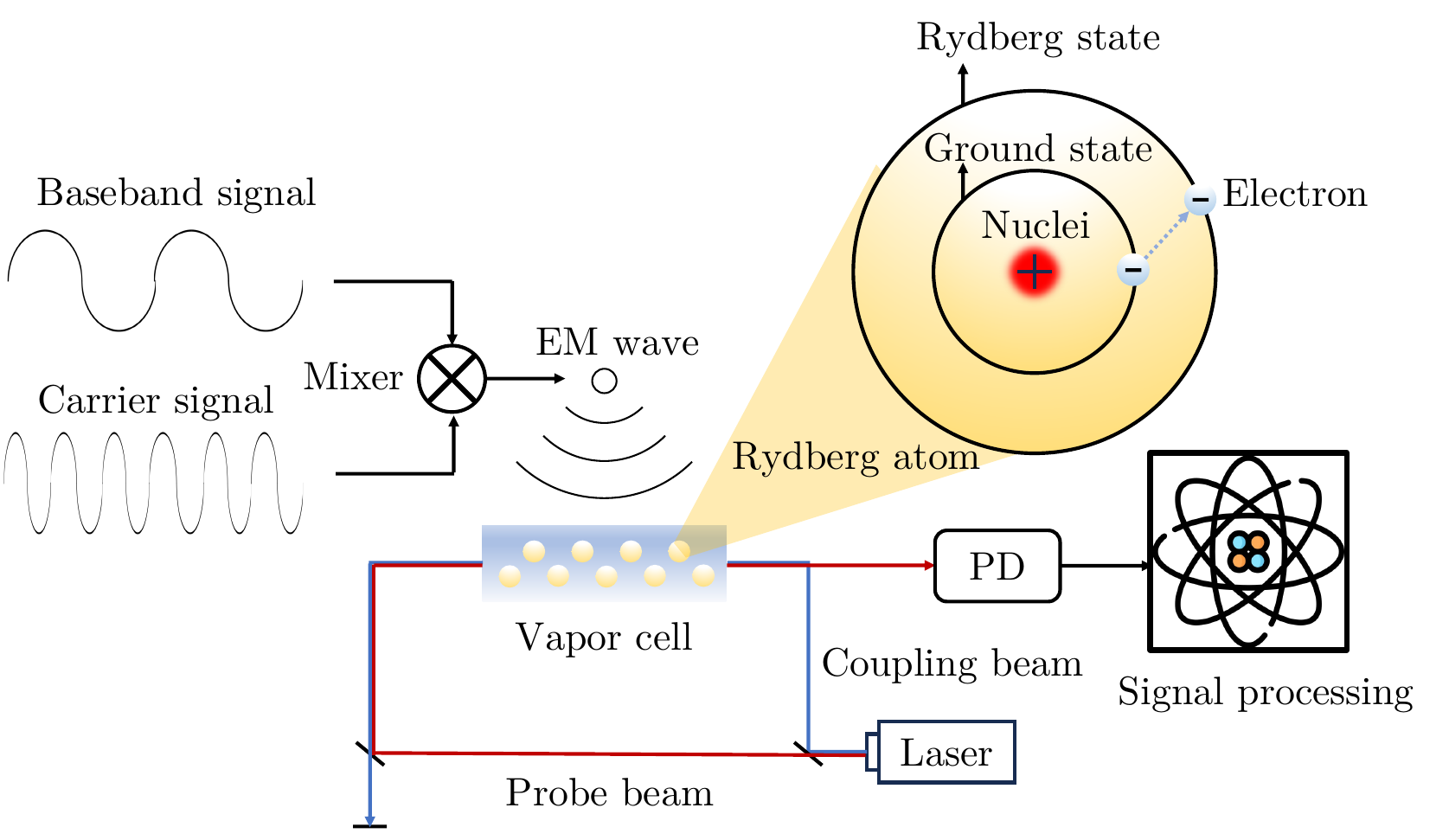}
        \caption{Wireless signal reception and processing based on an atomic receiver. When the EM wave is absorbed or emitted, the energy level of the electron is transferred to another energy level, which is known as atomic level transition \cite{AtomicTrans}.}
        \label{Fig1} 
    \end{center}
\end{figure}

While the incorporation of quantum wireless systems and communications is emerging \cite{Atomic-MIMO, IQ}, research for quantum wireless sensing is still at the nascent stage \cite{QuantumWirel, Rydberg_AoA1, Rydberg_AoA2}. For example, authors in \cite{Rydberg_AoA1} employed the simple geometric relationship between a phase difference and angle-of-arrival (AoA) of a single target, which is not applicable for multi-user environments. Although several hardware demonstrations have validated the effectiveness of quantum wireless sensing \cite{QuantumWirel, Rydberg_AoA2}, developing the signal processing algorithm for quantum wireless sensing has not been investigated yet. 

Motivated by this observation, in this paper, we propose a Quantum-MUSIC, multiple signal classification (MUSIC) for quantum wireless sensing. To the best of our knowledge, the proposed algorithm is the first signal processing algorithm for quantum wireless sensing of multi-user. Here, the received signal of the atomic receiver is not directly applicable to signal processing due to the magnitude-only measurement. Therefore, we modify the biased GS algorithm in \cite{Atomic-MIMO} to recover the quantum wireless channel to generate the covariance matrix. Thereafter, applying the MUSIC algorithm enables accurate quantum wireless sensing in the multi-user environment. In this paper, we compare the performance of traditional RF-based MUSIC and the proposed algorithm and present how the atomic receiver and proposed algorithm improve the sensing performance.

$\textit{Notations:}$ $(\cdot)^{\mathsf{T}}$ and $(\cdot)^{\mathsf{H}}$ respectively denote transpose and conjugate transpose. $(\cdot)^{-1}$, $(\cdot)^{*}$ denote the inverse, complex conjugate, and $\textrm{diag}(\cdot)$ denotes the diagonal matrix. $|\mathbf{\mathbf{\cdot}}|$ denotes the magnitude and $\lVert \cdot \rVert_{\textrm{2}}$ denotes L2 norm. $\circ$ denotes the Hadamard product. The notation $\mathcal{CN}(\boldsymbol{\mu},\sigma^2\mathbf{I})$ denotes a circularly-symmetric complex Gaussian distribution whose mean is $\boldsymbol{\mu}$ and covariance is $\sigma^2$.

\begin{figure}[t]
    \begin{center}
        \includegraphics[width=1\columnwidth,center]{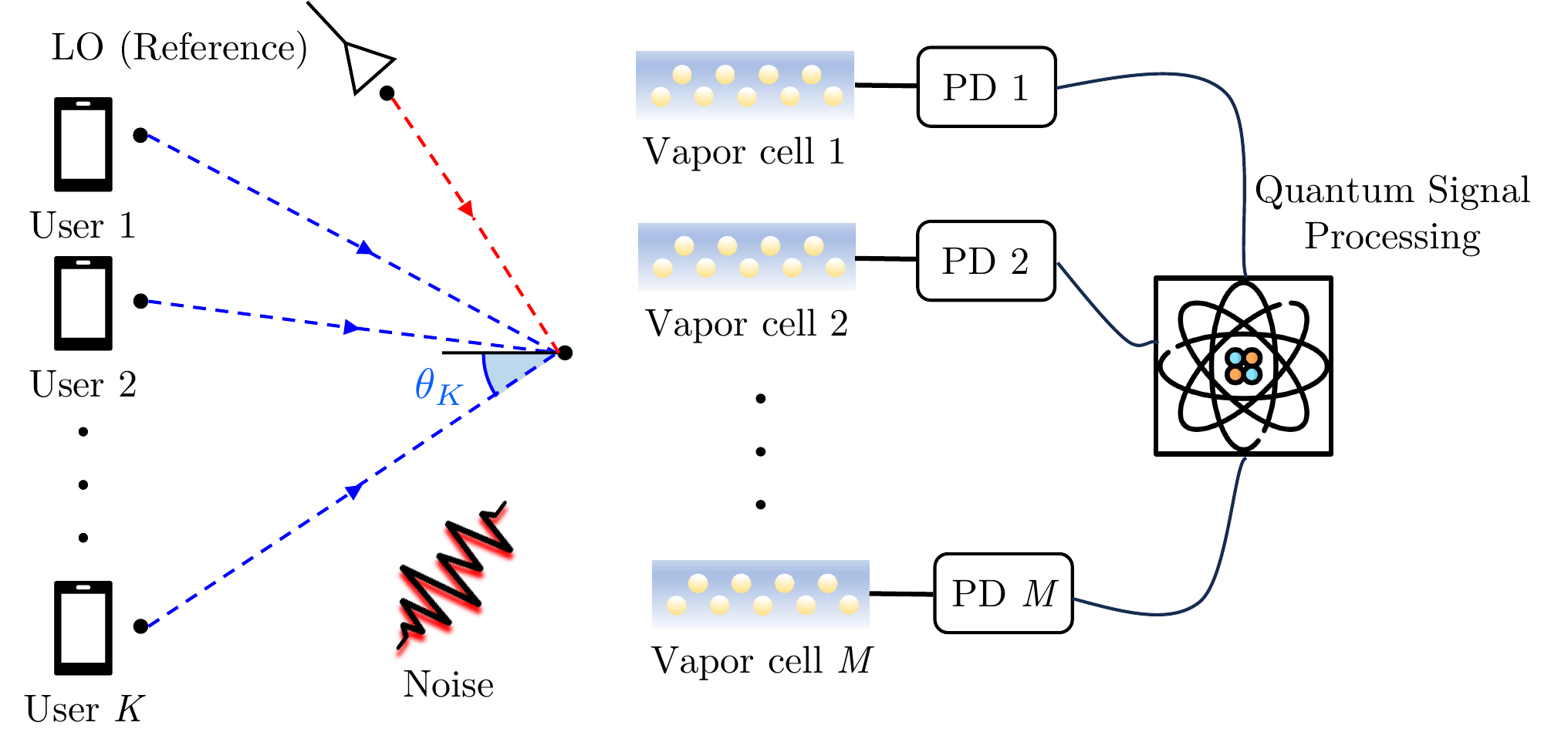}
        \caption{A scheme of an atomic array-based quantum signal processing system. Here, a number of $K$ users transmit the RF signal to an atomic array comprising $M$ vapor cell and the equivalent number of PDs.}
        \label{Fig2} 
    \end{center}
\end{figure}

\section{Quantum Mechanics and System Model for Quantum Wireless Sensing}\label{Sec2}

We consider the uplink narrowband multi-user scenario where $K$ users with a single antenna transmit the RF signal to the atomic receiver as depicted in Fig.~\ref{Fig2}. We considered the wireless channel consists of a single line-of-sight (LoS) path per user. The atomic receiver comprises $M$ vapor cells filled with Rydberg atoms, acting as atomic antennas for measurement acquisition, and an equivalent number of photodetectors (PD) with a single local oscillator (LO). 
\subsection{Quantum Mechanics for Measurement Acquisition}
The quantum state of Rydberg atoms transfers from the ground state to the excited state when the EM wave is absorbed by the electron of Rydberg atoms \cite{QuantumWirel}. Due to these two distinct energy levels, the quantum state of the Rydberg atom can be represented as $2 \times 1$ complex vector according to the two-state quantum system\footnote{In quantum mechanics, the Bra-Ket notation is used for the representation of quantum states \cite{Bra_Ket}. Here, $\bra{\Psi}$ and $\ket{\Psi}$ denote the "bra" and "ket", which represent the row vector and the column vector in the Hilbert space. The inner product and the outer product of two kets $\ket{\Psi}$ and $\ket{\varphi}$ are represented as $\langle\Psi \mid \varphi\rangle$, $|\Psi\rangle\langle \varphi|$, respectively.}.   Let $\ket{g}=[0,1]^{\mathsf{T}}$ and $\ket{e}=[1,0]^{\mathsf{T}}$ denote ground state and excited state, respectively. The general quantum state for each Rydberg atoms inside the $m$-th vapor cell, $\ket{\Psi}_{m}$, can be denoted as follows \cite{Atomic-MIMO}:
\begin{equation}
    \ket{\Psi}_{m}=\alpha_{g,m}\ket{g}+\alpha_{e,m}\ket{e}=[\alpha_{e,m},\alpha_{g,m}]^{\mathsf{T}},
\end{equation}
where $\alpha_{g,m}$ and $\alpha_{e,m}$ represent the probability amplitudes of the ground state and excited state inside $m$-th vapor cell, satisfying $\left| \alpha_{g,m} \right|^2+\left| \alpha_{e,m} \right|^2=1$. 

To monitor the state evolution of the quantum state, the equation (1) is considered as $\ket{\Psi(t)}_{m}=\alpha_{g,m}(t)\ket{g}+\alpha_{e,m}(t)\ket{e}$ in a time-varying system. Here, the state evolution of $\ket{\Psi(t)}_{m}$ is governed by the Schrödinger equation \cite{Quantum_Mech}:
\begin{equation}
    i \hbar \frac{\partial|\Psi(t)\rangle_m}{\partial t}=\left(\hat{H}_m+\hat{V}_m\right)|\Psi(t)\rangle_m ,
\end{equation}
where $\hat{H}_{m}$ and $\hat{V}_{m}$ represent the free Hamiltonian and interaction Hamiltonian operators, respectively. Here, the free Hamiltonian means the energy of an isolated atom. Since all users share the same angular frequency $\omega$, we assumed all atoms inside vapor cells are excited to uniform Rydberg states \cite{Atomic-MIMO}. The free Hamiltonian is expressed as $\hat{H}_{m}=\textrm{diag}(\hbar \omega_e, \hbar \omega_g)$, where $\omega_e$ and $\omega_g$ represent the angular frequency of excited state and ground state\footnote{The energy level of the ground state $E_g=\hbar \omega_g$ and the excited state $E_e=\hbar \omega_e$ can be acquired by acting $\hat{H}_{m}$ on $\ket{g}$ and $\ket{e}$. Thereby, the free Hamiltonian $\hat{H}_{m}$ is denoted as $2\times2$ diagonal matrix as $\hat{H}_{m}=\hbar \omega_e|e\rangle\left\langle e\left|+\hbar \omega_g\right| g\right\rangle\langle g|=\textrm{diag}(\hbar \omega_e, \hbar \omega_g)$ \cite{Atomic-MIMO}.}. $\hbar=\frac{h}{2 \pi}$ is the reduced Planck constant, where $h=6.6626 \times 10^{-34} \mathrm{J} \cdot \mathrm{s}$ is the Planck constant. 

Meanwhile, the interaction Hamiltonian $\hat{V}_{m}$, which represents the energy of interaction between EM wave and atom, is expressed as \cite{Atomic-MIMO}:

\begin{equation}
\hat{V}_{m}=\xi_m(t)|e\rangle\left\langle g\left|+\xi_m^*(t)\right| g\right\rangle\langle e|,
\end{equation}
 where $\xi_m(t)=\sum_{k=1}^K \boldsymbol{\mu}_{e g}^\mathsf{T} \boldsymbol{\epsilon}_{m, k} \sqrt{P_k} \rho_{m, k}s_k \cos \left(\omega t+\varphi_{m, k}\right)$. Here, $\boldsymbol{\mu}_{e g}$ is the transition dipole moment \cite{Quantum2}. Note that EM wave at $m$-th vapor cell $\mathbf{E}_{m}(t)=\sum_{k=1}^K \boldsymbol{\epsilon}_{m, k} \sqrt{P_k} \rho_{m, k}s_k \cos \left(\omega t+\varphi_{m, k}\right)$ is encoded in $\xi_m(t)$, where $\boldsymbol{\epsilon}_{m, k}, \rho_{m, k},$ and $\varphi_{m, k}$ represent the polarization direction, path loss, and phase shift of the EM wave propagation from the $k$-th user to $m$-th vapor cell. We considered the uniform linear array (ULA) based receiver so that the channel coefficient $\rho_{m, k}e^{j\varphi_{m,k}}$ is represented as $\rho_{m,k} e^{j \varphi_{m,k}}=\alpha e^{j {m d \sin \theta_{k}}/{\lambda}}$, where $\theta_{k}$ and $\alpha$ are the AoA from $k$-th user and the channel gain. $P_k, s_{k}$ represent the power and the transmitted signal from $k$-th user, respectively.
 Eventually, replacing $\hat{H}_{m}$, $\hat{V}_{m}$ into (2) and applying the rotating wave approximation \cite{quantum_rotating} derives the probability amplitude of excited state at $m$-th vapor cell $\left|\alpha_{e, m}(t)\right|^2$ as \cite{Atomic-MIMO}:
 \begin{equation}
    \left|\alpha_{e, m}(t)\right|^2=\sin ^2\left(\frac{\Omega_{m}}{2} t\right),
\end{equation}
 where the Rabi frequency at $m$-th vapor cell $\Omega_{m}$ is given as
 \begin{equation}
    \Omega_{m}=\left|\sum_{k=1}^K \frac{1}{\hbar}\left(\boldsymbol{\mu}_{e g}^\mathsf{H} \boldsymbol{\epsilon}_{m, k}\right) \sqrt{P_k} \rho_{m, k}  s_k e^{j\varphi_{m, k}}\right|.
\end{equation}

\begin{figure}[t]
    \begin{center}
        \includegraphics[width=0.95\columnwidth,center]{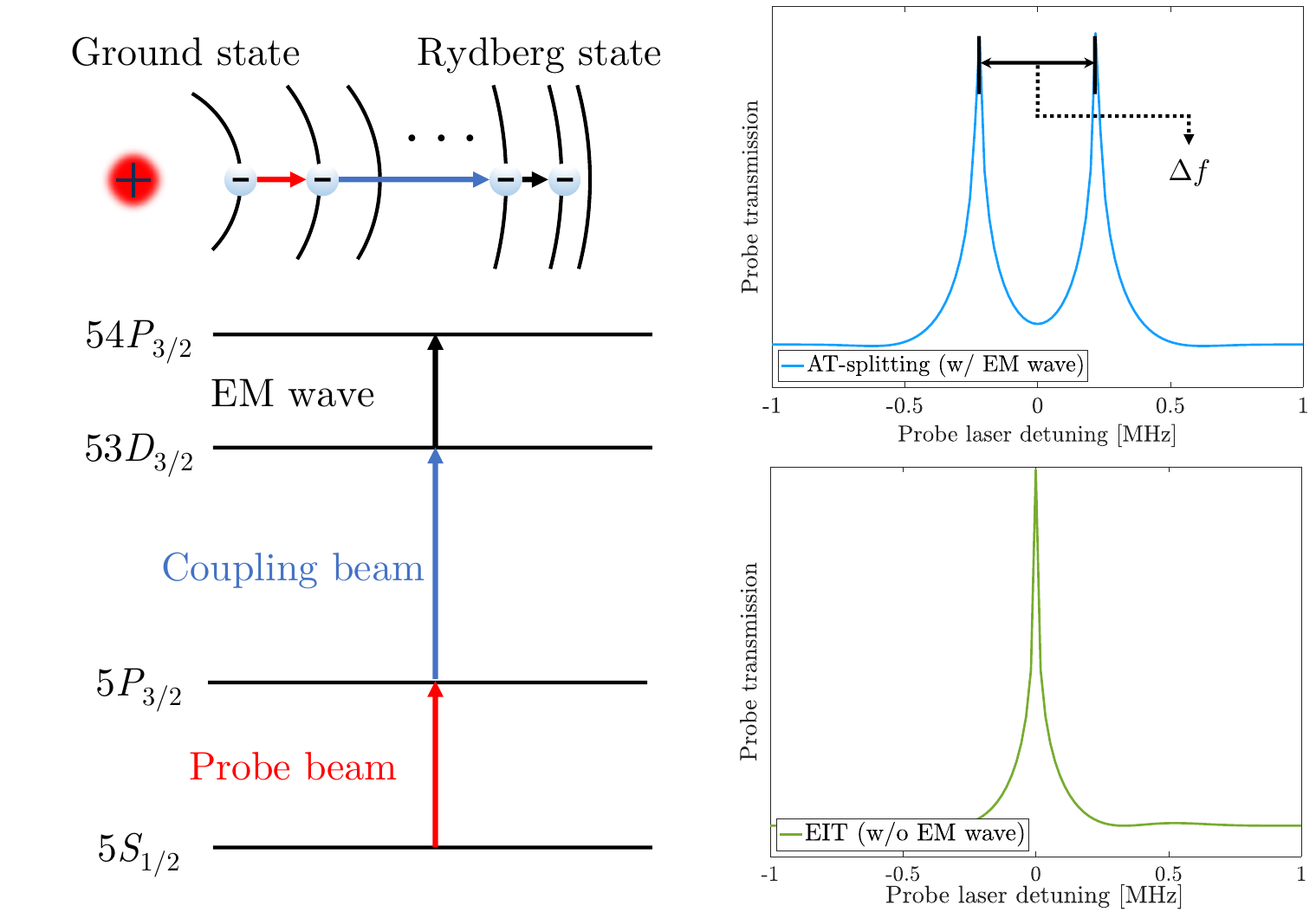}
        \caption{Example of measuring the Rabi frequency via EIT and AT-splitting \cite{EIT-AT}. The probe beam passes the vapor cell without absorption, causing the EIT phenomenon. Thereafter, the two Rydberg states (highly excited quantum states which are $53D_{3/2}$ and $54P_{3/2}$) are coupled by the EM wave, which splits the spectrum of the probe light, bringing the AT-splitting effect.}
        \label{Fig3} 
    \end{center}
\end{figure}

The Rabi frequency at $m$-th vapor cell $\Omega_{m}$ can be inferred by two quantum phenomena, which are electromagnetically induced transparency (EIT) and Autler-Townes (AT) splitting as shown in Fig. \ref{Fig3} \cite{EIT-AT}. Here, measuring the splitting interval $\Delta f$ via PD allows the inducing of Rabi frequency, which is equivalent to the measurement of the received signal.
\subsection{Atomic Receiver System Model}
  Denote the measurement of Rabi frequency at $m$-th atomic antenna $\hat{\Omega}_{m}$ as $\hat{\Omega}_{m}=y_{m}$. Then, the received signal $\mathbf{y}=[{y}_{1}, {y}_{2}, \cdots, {y}_{M}]^{\mathsf{T}}$ can be expressed as follows
\begin{equation}
    \mathbf{y}=\left|\mathbf{A}^\mathsf{H} \mathbf{s}+\mathbf{b}+\mathbf{n}\right| \in \mathbb{R}^{M \times 1},
\end{equation}
where $\mathbf{A}=[\mathbf{a}_{1}, \mathbf{a}_{2}, \cdots,\mathbf{a}_{M}] \in \mathbb{C}^{K \times M}$, $\mathbf{b}=\left[b_1, \cdots, b_M\right]^\mathsf{T} \in \mathbb{C}^{M \times 1}$, and $\mathbf{s}=[s_{1}, s_{2}, \cdots,s_{K}]^{\mathsf{T}} \in \mathbb{C}^{K \times 1}$ denote the wireless channel of atomic receiver, reference signal, and the transmitted signal from users with single snapshot, respectively. Here, the channel vector at $m$-th atomic antenna $\mathbf{a}_{m}$ is denoted as $\mathbf{a}_{m}=[a_{m,1}, a_{m,2}, \cdots, a_{m,K}]^{\mathsf{T}} \in \mathbb{C}^{K \times 1}$, where $a_{m, k}=\frac{1}{\hbar}\left(\boldsymbol{\mu}_{e g}^\mathsf{T} \boldsymbol{\epsilon}_{m, k}\right) \sqrt{P_k} \rho_{m, k} e^{-j \varphi_{m, k}}$. 
Note that we adopted the holographic phase sensing scheme to remove the phase ambiguity \cite{Atomic-MIMO}, where $b_m=\frac{s_b}{\hbar} \boldsymbol{\mu}_{e g}^\mathsf{H} \boldsymbol{\epsilon}_{m, b} \sqrt{P_b} \rho_{m, b} e^{j \varphi_{m, b}}$. Here, $s_b, \rho_{m, b}, \varphi_{m, b},$ and $\varphi_{m, b}$ are known reference signal, path loss, phase shift, and polarization direction of the LO to atomic receiver path, respectively. Last, the measurement noise is modeled as the Gaussian distribution, $\mathbf{n} \sim \mathcal{C N}\left(\boldsymbol{0}, \sigma_n^2 \mathbf{I}\right)$, where $\sigma_n^2$ denotes the quantum shot noise (QSN) power of the atomic receiver. 

\section{Multi-USER Quantum Wireless Sensing Algorithm}
 In this section, we introduce Quantum-MUSIC for the quantum wireless sensing of multi-user. Since the received signal is in magnitude form in (6), retrieving the channel information is required. Recovering the channel information via atomic receiver has an identical mathematical form as retrieving the phase from the magnitude-only measurement, which is well-known as the PR problem \cite{Phase_Retrieval}. To be specific, due to the magnitude-only measurements, the received signal satisfies the following equation ${\mathbf{y}}=\left|\mathbf{A}^\mathsf{H} \mathbf{s}+\mathbf{b}+\mathbf{n}\right|=\left|\mathbf{s}^\mathsf{H} \mathbf{A}+\mathbf{b}^{*}+\mathbf{n}^*\right|$. Owing to this structure, recovering the unknown channel can be achieved in the same way as solving the PR problem. 
 
Let the pilot signal be transmitted a total of $P$ times so that transmitted signal for single snapshot $\mathbf{s}$ is replaced to pilot signal matrix $\mathbf{S}$, where $\mathbf{S} = [\mathbf{s}_{1}, \mathbf{s}_{2}, \cdots, \mathbf{s}_{P}] \in \mathbb{C}^{K \times P}$. Then, the received signal at $m$-th atomic antenna $\mathbf{z}_{m}=[{z}_{m,1}, \cdots, {z}_{m, P}]^{\mathsf{T}} \in \mathbb{R}^{P \times 1}$ is represented as
\begin{equation}
{\mathbf{z}}_{m}=\left|\mathbf{S}^\mathsf{H}\mathbf{a}_{m}+\mathbf{b}_{m}+\mathbf{n}_{m}\right|,
\end{equation}
where $\mathbf{b}_m=\left[b_{m}, \cdots, b_{m}\right]^\mathsf{H} \in \mathbb{C}^{P \times 1}$ and $\mathbf{n}_{m} \sim \mathcal{C N}\left(\boldsymbol{0}, \sigma_n^2 \mathbf{I}\right)$ are the reference signal, QSN at the $m$-th atomic antenna. Since the phase of $\mathbf{z}_{m}$, $e^{j\angle\mathbf{z}_{m}}=[e^{j\angle{z}_{m,1}},e^{j\angle{z}_{m,2}},\cdots,e^{j\angle{z}_{m, P}}]^{\mathsf{T}}$, is inaccessible, assume ${\bar{\mathbf{z}}}_{m}=\mathbf{S}^\mathsf{H}\mathbf{a}_{m}+\mathbf{b}_{m}+\mathbf{n}_{m}=[\bar{z}_{m,1},\bar{z}_{m,2},\cdots,\bar{z}_{m,P}]^{\mathsf{T}}$. Then, recovering the channel can be formulated as the joint optimization problem of $e^{j\angle\mathbf{\bar{z}}_{m}}$ and $\mathbf{a}_{m}$, which is given as 
\begin{equation}
    \min _{\mathbf{a}_{m},\angle\mathbf{\bar{z}}_{m}}\left\|\mathbf{z}_{m} \circ e^{j \angle{\mathbf{\bar{z}}}_{m}}-\mathbf{S}^\mathsf{H}\mathbf{a}_{m}-\mathbf{b}_{m}\right\|_2^2,
    \label{eq} 
\end{equation}
where $e^{j \angle{\mathbf{\bar{z}}}_{m}}=[e^{j \angle\bar{z}_{m,1}}, e^{j \angle\bar{z}_{m,2}}, \cdots, e^{j \angle\bar{z}_{m, P}}]^{\mathsf{T}}$. In the following sub-section, we introduce the channel recovery method by solving the equation (\ref{eq}).

\begin{algorithm}[t]
  \caption{Quantum-MUSIC}
  \Input{$\mathbf{S}$, $\mathbf{b}_{m}$, $N$, $P$, $\mathbf{z}_{m}$ for $m=1,\cdots,M$}

  \For{$m={1,2,\cdots,M}$}{

  Generate the expanded signal matrix $\bar{\mathbf{S}}=\left[\mathbf{S}^\mathsf{H}, \mathbf{b_{m}}\right]^\mathsf{H}=\left[\bar{\mathbf{s}}_1, \bar{\mathbf{s}}_2, \cdots \bar{\mathbf{s}}_P\right]$
    
  Find the principal eigenvector $\mathbf{v}$ of $\bar{\mathbf{R}}=\sum_{p=1}^P z_{m,p} {\bar{\mathbf{s}}}_p {\bar{\mathbf{s}}}_p^\mathsf{H}$
  
   Set $\bar{r}=\frac{|\mathbf{v}^\mathsf{H} \bar{\mathbf{S}}| \mathbf{z}_{m}}{\|\bar{\mathbf{S}}^\mathsf{H} \mathbf{v}\|_2^2},$ and $\mathbf{\bar{a}}_{m}^{0}=\bar{r} \mathbf{v}$

  Initialize $\mathbf{a}_{m}^{0}=\mathbf{\bar{a}}_{m}^{0}(1: K)$
  
  \For{$n={1,2,\cdots,N}$}{
  Update $\angle {\mathbf{\bar{z}}}_m^n=\angle\left(\mathbf{S}^{\mathsf{H}} \mathbf{a}_m^{n-1}+\mathbf{b}_{m}\right)$
  
  Update $\mathbf{a}_m^n=(\mathbf{S S}^{\mathsf{H}})^{-1} \mathbf{S}(\mathbf{z}_m \circ e^{j\angle{\mathbf{\bar{z}}}_m^n}-\mathbf{b}_{m}),$

  }

  $\hat{\mathbf{a}}_{m}={\mathbf{a}}_{m}^{N}$}
  
  Estimated channel matrix $\hat{\mathbf{A}}=[\hat{\mathbf{a}}_{1}, \hat{\mathbf{a}}_{2}, \cdots,\hat{\mathbf{a}}_{M}]^{\mathsf{H}}$ 

  Find the noise subspace $\mathbf{U}_{\mathbf{N}}$ of  $\mathbf{R}=\frac{1}{P}\hat{\mathbf{A}}\hat{\mathbf{A}}^{\mathsf{H}}$
  
  Construct the Quantum-MUSIC spectrum $\mathbf{P}_{\mathbf{Q}}(\theta)=\frac{1}{\boldsymbol{a}^{\mathsf{H}}(\theta) \mathbf{U}_{\mathbf{N}}\mathbf{U}_{\mathbf{N}}^{\mathsf{H}} \boldsymbol{a}(\theta)}$

Find $\left\{\hat{\theta}_k\right\}_{k=1}^K$ from the $K$ highest peaks of $\mathbf{P}_{\mathbf{Q}}(\theta)$

\Output{The estimated channel parameter $\left\{\hat{\theta}_k\right\}_{k=1}^K$}
\end{algorithm}

\subsection{Channel Recovery from Magnitude-only Measurement}
To recover the quantum wireless channel $\mathbf{A}$, we modified the biased GS algorithm in \cite {Atomic-MIMO}. To begin with, we generated an expanded signal matrix $\bar{\mathbf{S}}=[\mathbf{S}^\mathsf{H}, \mathbf{b_{m}}]^\mathsf{H}=[\bar{\mathbf{s}}_{1}, \cdots, \bar{\mathbf{s}}_{P}] \in \mathbb{C}^{(K+1) \times P}$ and expanded channel vector $\bar{\mathbf{a}}_{m}=[{\mathbf{a}}_{m}^\mathsf{H}, 1]^\mathsf{H} \in \mathbb{C}^{(K+1) \times 1}$. Then, the received signal at $m$-th atomic array element can be expressed as typical PR measurement form, which is represented as $\mathbf{z}_{m}=\left|\mathbf{S}^\mathsf{H}\mathbf{a}_{m}+\mathbf{b}_{m}+\mathbf{n}_{m}\right|=\left|\bar{\mathbf{S}}^\mathsf{H} \bar{\mathbf{a}}_{m}+\mathbf{n}_{m}\right|$. This process allows the employing of the spectral method-based initialization \cite{Wirtinger}, which is important since the optimization fails with an inaccurate initial point. The initialization process starts with constructing the expanded covariance matrix $\bar{\mathbf{R}}$, which is represented as
\begin{equation}
    \bar{\mathbf{R}}=\sum_{p=1}^P z_{m,p} {\bar{\mathbf{s}}}_p {\bar{\mathbf{s}}}_p^\mathsf{H} \in \mathbb{C}^{(K+1) \times (K+1)}.
\end{equation}
Thereafter, the initial value $\mathbf{\bar{a}}_{m}^{0}$ of $\mathbf{\bar{a}}_{m}$ is calculated as $\mathbf{\bar{a}}_{m}^{0}=\bar{r} \mathbf{v}$ where $\mathbf{v}$ is the principal eigenvector of $\bar{\mathbf{R}}$ and $\bar{r}$ is the magnitude of $\mathbf{\bar{a}}_{m}^{0}$, which is represented as 
\begin{equation}
   \bar{r}=\frac{|\mathbf{v}^\mathsf{H} \bar{\mathbf{S}}| \mathbf{z}_{m}}{\|\bar{\mathbf{S}}^\mathsf{H} \mathbf{v}\|_2^2}.
\end{equation}
Since the $K$ elements of $\mathbf{\bar{a}}_{m}^{0}$ are channel vectors from users, $\mathbf{a}_{m}^{0}$ is initialized by taking $K$ elements from $\mathbf{\bar{a}}_{m}^{0}$, which is represented as $\mathbf{a}_{m}^{0}=\mathbf{\bar{a}}_{m}^{0}(1: K)$, completing the initialization process. After the initialization, the joint optimization is carried out, which is represented as
\begin{equation}
    \angle {\mathbf{\bar{z}}}_m^n=\angle(\mathbf{S}^{\mathsf{H}} \mathbf{a}_m^{n-1}+\mathbf{b}_{m}),
\end{equation}
\begin{equation}
    \mathbf{a}_m^n=(\mathbf{S S}^{\mathsf{H}})^{-1} \mathbf{S}(\mathbf{z}_m \circ e^{j\angle{\mathbf{\bar{z}}}_m^n}-\mathbf{b}_{m}),
\end{equation}
After the $N$ iteration, the estimated channel of $m$-th atomic array element is derived as $\hat{\mathbf{a}}_{m}={\mathbf{a}}_{m}^{N} \in \mathbb{C}^{M \times 1}$. Eventually, the estimated channel matrix is obtained by solving the optimization problem (\ref{eq}) on all atomic antenna elements, which is represented as $\hat{\mathbf{A}}=[\hat{\mathbf{a}}_{1}, \hat{\mathbf{a}}_{2}, \cdots,\hat{\mathbf{a}}_{M}]^{\mathsf{H}} \in \mathbb{C}^{M \times K}$.
\subsection{Multi-User Parameter Estimation}
We integrated a MUSIC \cite{MUSIC} into our proposed algorithm for the sensing of multi-user. MUSIC is a subspace-based algorithm exploiting the orthogonality between noise subspace and the steering vectors of multi-user. The ULA far-field steering vector $\boldsymbol{a}(\theta) \in \mathbb{C}^{M \times 1}$ in terms of AoA $\theta$ is expressed as
 \begin{equation}
    \boldsymbol{a}(\theta)=\frac{1}{\sqrt{M}}\left[1, e^{-j\frac{2\pi d \sin\theta}{\lambda}}, \cdots, e^{-j(M-1)\frac{2\pi d \sin\theta}{\lambda}}\right]^{\mathsf{T}},
\end{equation}
where $d$ and $\lambda$ denote the distance between adjacent array elements and the wavelength of the received RF signal, respectively. After the channel is recovered, the channel covariance matrix is derived as $\mathbf{R}=\frac{1}{P}\hat{\mathbf{A}}\hat{\mathbf{A}}^{\mathsf{H}}\in \mathbb{C}^{M \times M}$. Thereafter, eigenvalue decomposition of $\mathbf{R}$ yields $\mathbf{R}=\mathbf{U} \mathbf{\Sigma} \mathbf{U}^{\mathsf{H}}$, where $\mathbf{U}\in \mathbb{C}^{M \times M}$ and $\mathbf{\Sigma}\in \mathbb{C}^{M \times M}$ denotes the eigenvectors and eigenvalues of $\mathbf{R}$ in descending order. Note that $\mathbf{U}$ can be decomposed as $\mathbf{U}=[\mathbf{U}_{\mathbf{S}}  \mathbf{U}_{\mathbf{N}}]$ according to the eigenvalues, where $\mathbf{U}_{\mathbf{S}} \in \mathbb{C}^{M \times K}$ and $\mathbf{U}_{\mathbf{N}} \in \mathbb{C}^{M \times (M-K)}$ denotes signal subspace and noise subspace, respectively. Then, the Quantum-MUSIC spectrum $\mathbf{P}_{\mathbf{Q}}(\theta)$ is defined as
 \begin{equation}
    \mathbf{P}_{\mathbf{Q}}(\theta)=\frac{1}{\boldsymbol{a}^{\mathsf{H}}(\theta) \mathbf{U}_{\mathbf{N}}\mathbf{U}_{\mathbf{N}}^{\mathsf{H}} \boldsymbol{a}(\theta)}.
\end{equation}
Here, the estimated peak values of the Quantum-MUSIC spectrum, $\left\{\hat{\theta}_k\right\}_{k=1}^K$, correspond to the estimated AoAs of multi-user.

\section{Simulation Results}

\subsection{Simulation Environments}
In the simulation, the number of atomic array elements, $M$, is 32. The number of pilot snapshots, $P$, is set to 100. The Rydberg energy levels $52D_{5/2}$ and $53P_{3/2}$ are exploited to detect the RF signals of the angular transition frequency $ \omega_{eg} \approx 2\pi \times 5$ GHz. By leveraging the ARC package \cite{Arc}, the transition dipole moment $\boldsymbol{\mu}_{e g}$ of $52D_{5/2}$ and $53P_{3/2}$ is obtained as $\left[0,1785.9 q a_0, 0\right]^T$, where $q=1.602 \times 10^{-19} \mathrm{C}$ is the charge of an electron and $a_0=5.292 \times 10^{-11} \mathrm{~m}$ is the Bohr radius. Considering the randomness of polarization direction, $\boldsymbol{\epsilon}_{m, k}, \boldsymbol{\epsilon}_{m, b}$ follow $\mathcal{N}\left(0, \frac{1}{3}\right)$. The number of iterations, $N$, is set to 50. The signal power of $k$-th user $P_k$ is assumed to be uniform for all users so that $P_k=\frac{1}{K}\sigma_s^2$ where $\sigma_s^2$ is the total transmitted signal power. The root mean square error (RMSE) is derived as 
 \begin{equation}
    \operatorname{RMSE}=\sqrt{\frac{1}{Q K} \sum_{q=1}^{Q} \sum_{k=1}^K\left|\hat{\theta}_{q, k}-\theta_{q,k}\right|^2}.
\end{equation}
Here, $Q$ denotes the number of Monte Carlo trials. For the RMSE analysis, Monte Carlo trials are conducted 2,000 times for every simulation.

\begin{figure}[t]
    \begin{center}
        \includegraphics[width=1\columnwidth,center]{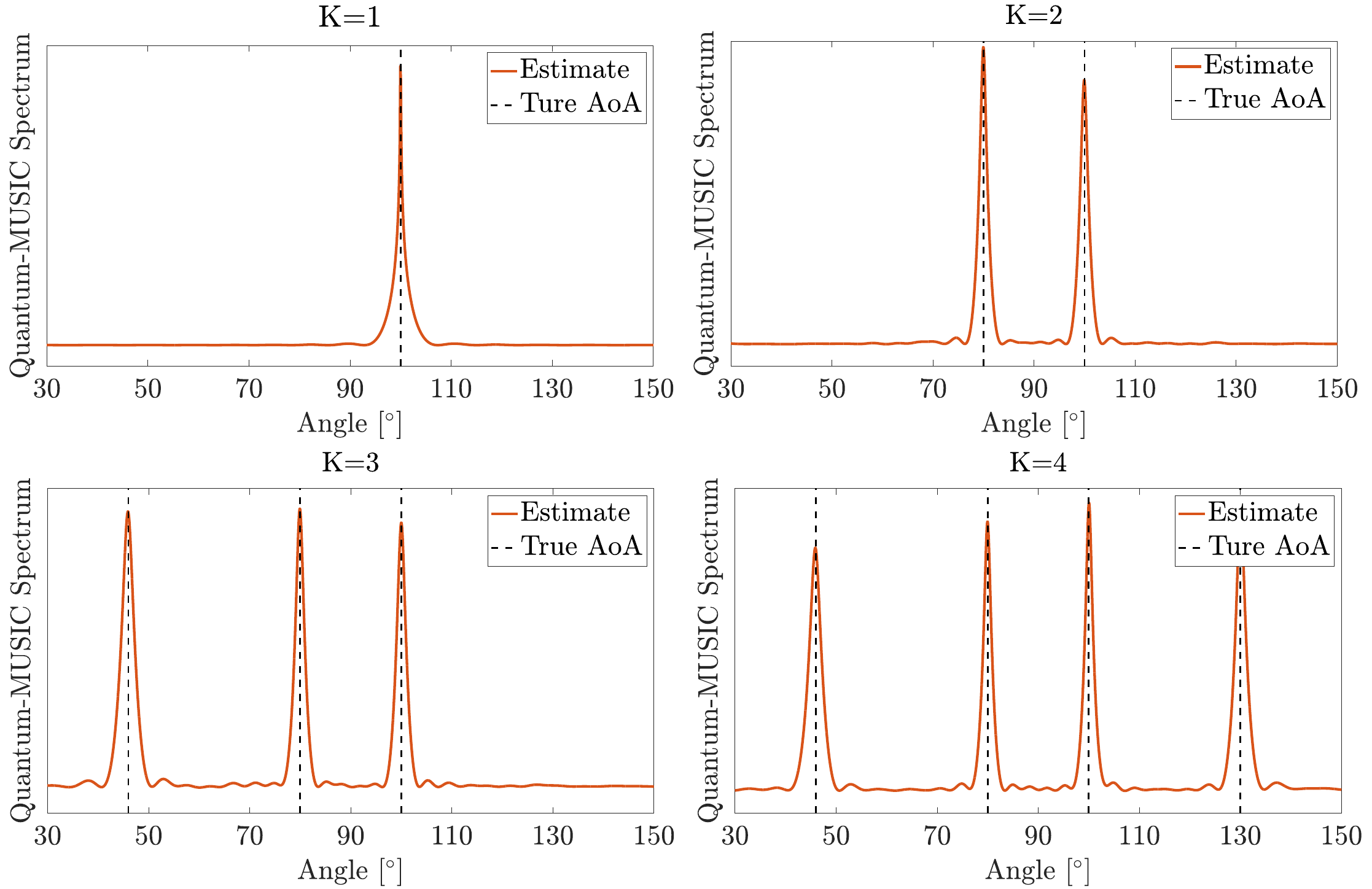}
        \caption{The Quantum-MUSIC spectrum according to the number of users. Here, the number of users, $K$, is set to $K=1, 2, 3, 4$.}
        \label{Fig4} 
    \end{center}
\end{figure}

\subsection{Analysis on Multi-User Sensing Performance}
We considered the AoA as a channel parameter for the analysis of multi-user sensing performance. Here, the angular grid is divided into $2^{14}$ for the calculation of $a(\theta)$ and the $\theta$ is uniformly distributed in the range of $[30^{\circ}, 150^{\circ}]$. We compared the performance of the proposed algorithm to the classical MUSIC \cite{MUSIC} algorithm in the RF domain to validate the potential of quantum wireless sensing. To model the RF antenna-based receiver system, the power density of Johnson-Nyquist thermal noise (JNTN) $\sigma_t^2$ is calculated \cite{QSN}.  

The Quantum-MUSIC spectrum according to the different number of users is shown in Fig. ~\ref{Fig4}. Here, the SNR is set to 10 dB. The simulation results show that the proposed algorithm can estimate the AoAs of different numbers of signal sources with magnitude-only measurement of the atomic receiver, overcoming the limitation of single-user-only estimation in the previous work \cite{Rydberg_AoA1}. 

\begin{figure}[t]
    \begin{center}
        \includegraphics[width=1.1\columnwidth,center]{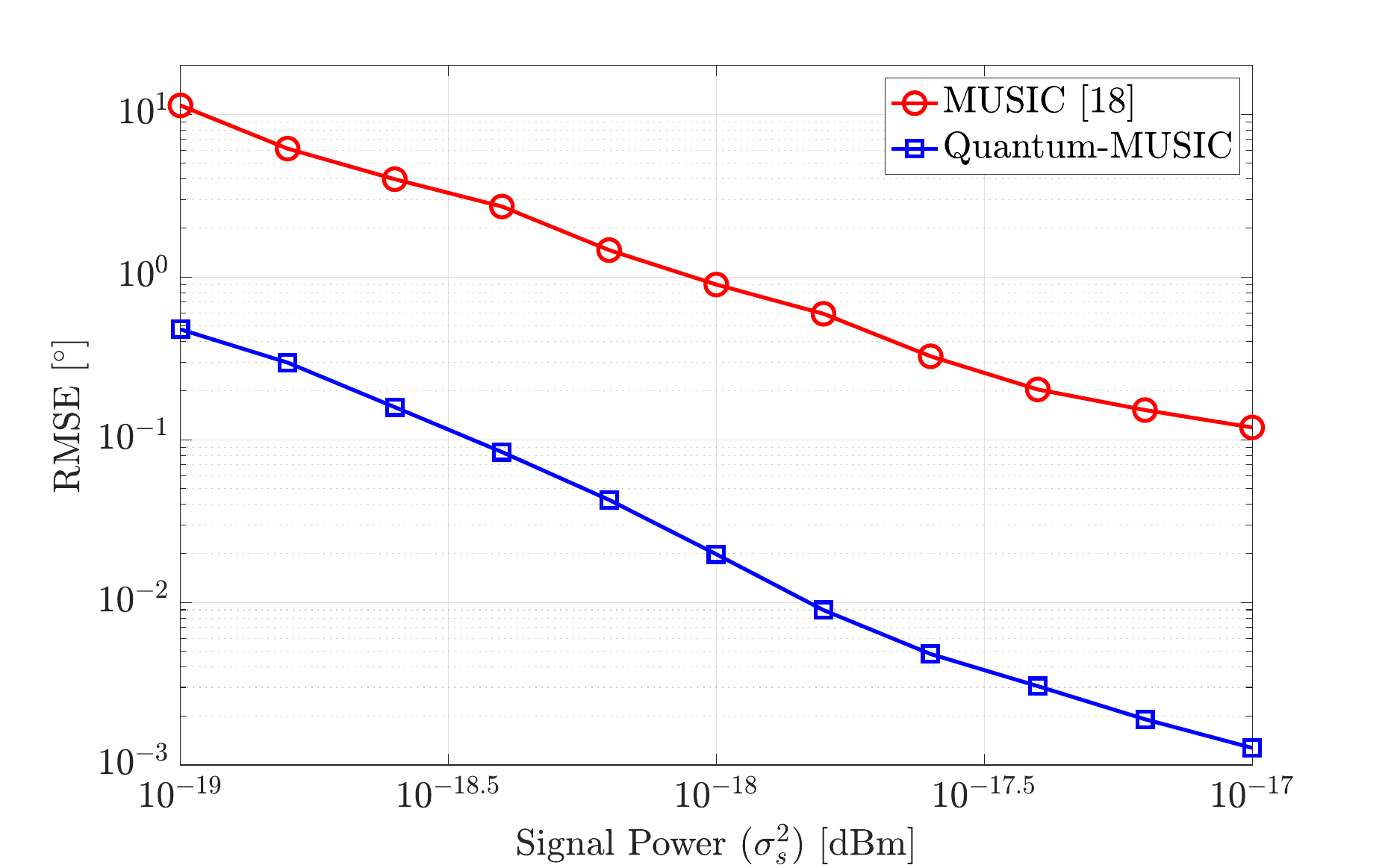}
        \caption{AoA RMSE versus signal power. Here, we compare the sensing performance where the number of users, $K$, is set as $K=3$.}
        \label{Fig5} 
    \end{center}
\end{figure}

The AoA RMSE according to signal power $\sigma_{s}^2$ is shown in Fig. ~\ref{Fig5}. Note that the signal-to-noise ratio (SNR) of the Quantum-MUSIC is generally different from that of MUSIC \cite{MUSIC} due to the difference between the power of QSN $\sigma_n^2$ and JNTN $\sigma_t^2$, making the fair comparison between quantum and RF wireless sensing challenging. To overcome this limitation, we first assumed that the transmitted signal power $\sigma_s^2$ is equal for both RF and quantum domains. Then, we calculated the noise power of two different domains, $\sigma_n^2$ and $\sigma_t^2$, for the room-temperature condition. Adopting the parameters in \cite{Atomic-MIMO} and \cite{QSN}, the power of QSN and JNTN are calculated as -191 dBm and -176 dBm, respectively, so that $\sigma_n^2=10^{-19.1}, \sigma_t^2=10^{-17.6}$ when $\omega=2 \pi \times 5 \mathrm{GHz}$. Thereafter, the transmitted signal power $\sigma_{s}^2$ is adjusted, enabling a fair comparison by the unified transmitted signal power. The simulation results show that the sensing accuracy of Quantum-MUSIC is superior to the MUSIC \cite{MUSIC} in all $\sigma_s^2$, implying that immunity to the thermal noise of atomic receiver is highly advantageous for wireless sensing of multi-user.   


\begin{figure}[t]
    \begin{center}
        \includegraphics[width=1.1\columnwidth,center]{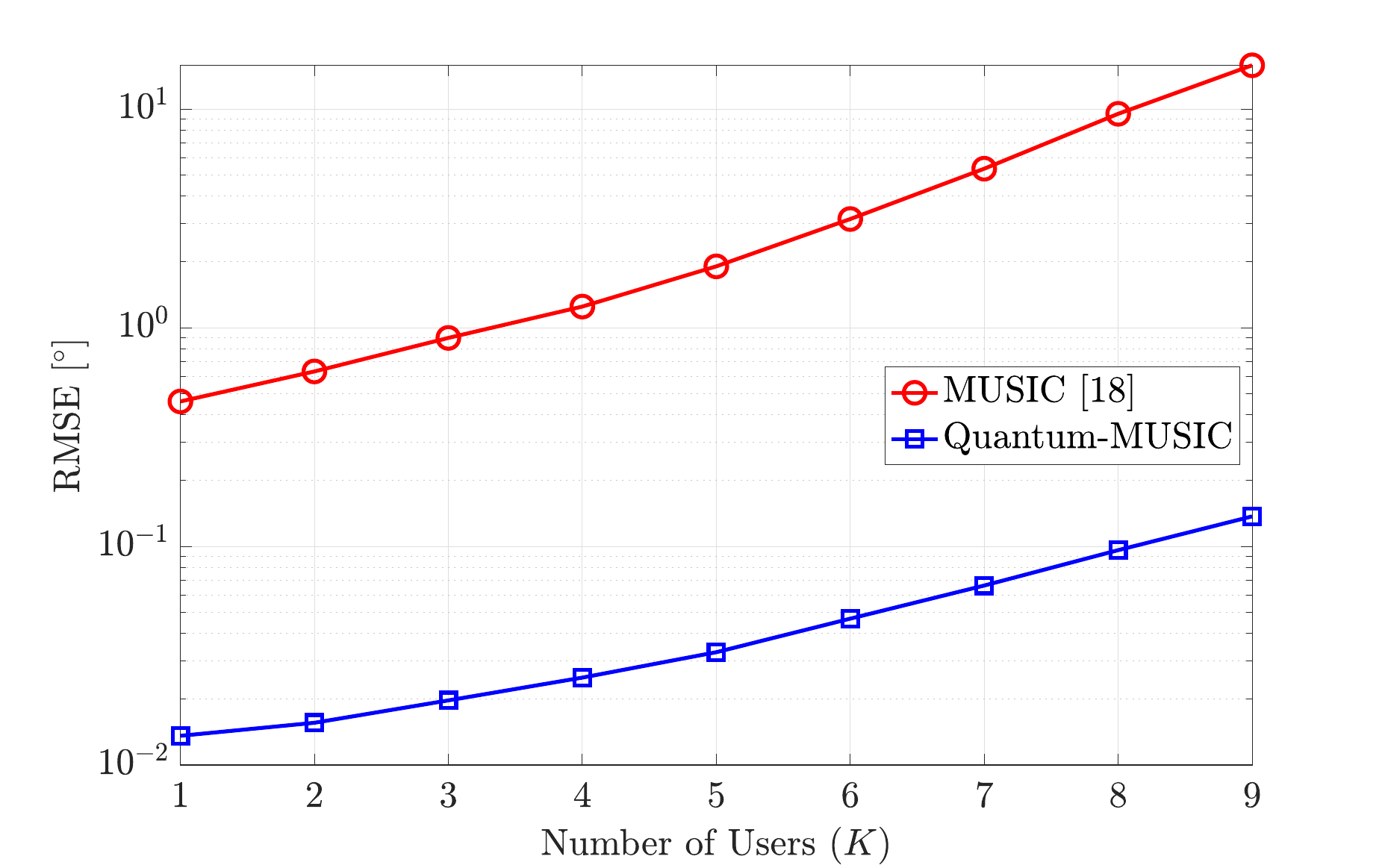}
        \caption{AoA RMSE versus the number of users. Here, the signal power, $\sigma_s^2$, is set to $\sigma_s^2=10^{-18}$.}
        \label{Fig6} 
    \end{center}
\end{figure}

To analyze the sensing performance versus the number of users, the AoA RMSE according to the number of users is shown in Fig. ~\ref{Fig6}. Here, we adopted the same noise power condition of both quantum and RF receiver in Fig. ~\ref{Fig5}, where $\sigma_n^2=10^{-19.1}$, $\sigma_t^2=10^{-17.6}$. The signal power of transmitted signal $\sigma_s^2$ is fixed to -180 dBm so that $\sigma_s^2=10^{-18}$. The proposed algorithm shows better performance compared to the MUSIC \cite{MUSIC} regardless of the number of users. The performance degradation of MUSIC \cite{MUSIC} compared to the Quantum-MUSIC increases along with the number of users, drawing the conclusion that the effectiveness of the low noise power of QSN increases as the number of user increases. 

\section{Conclusions}
In this paper, we propose Quantum-MUSIC for quantum wireless sensing of multi-user. The proposed algorithm first recovers the channel from the magnitude-only measurement by solving the joint optimization problem. Thereafter, the principle of traditional MUSIC is adopted for multi-user sensing. Simulation results show that the proposed algorithm outperforms MUSIC, implying that the minor noise level of QSN brings a high advantage for accurate wireless sensing. Future work is expected to develop a signal processing algorithm for multi-band quantum wireless sensing, where multi-user transmit an extremely broad range of frequencies from megahertz to terahertz.

\ifCLASSOPTIONcaptionsoff
  \newpage
\fi

\bibliographystyle{ieeetr}
\bibliography{reference}

\begin{thebibliography}{10}

\bibitem{QuantumWirel}
F.~{Zhang}, B.~{Jin}, Z.~{Lan}, Z.~{Chang}, D.~{Zhang}, Y.~{Jiao}, M.~{Shi}, and J.~{Xiong}, ``Quantum wireless sensing: Principle, design and implementation,'' in {\em Proc. the 29th Annu. Int. Conf. Mob. Comput. Netw. (ACM Mobicom'23)}, pp.~1--15, Jun. 2023.

\bibitem{atomic_receiver}
D.~A. Anderson, R.~E. Sapiro, and G.~Raithel, ``An atomic receiver for {AM} and {FM} radio communication,'' {\em IEEE Trans. Antennas and Propag.}, vol.~69, pp.~2455--2462, Apr. 2021.

\bibitem{Quantum_Phase}
D.~A. {Anderson}, R.~E. {Sapiro}, and G.~{Raithel}, ``Rydberg atoms for radio-frequency communications and sensing: Atomic receivers for pulsed {RF} field and phase detection,'' {\em IEEE Aerosp. Electron. Syst. Mag.}, vol.~35, pp.~48--56, Apr. 2020.

\bibitem{Rydberg_Atoms1}
C.~T. {Fancher}, D.~R. {Scherer}, M.~C.~S. {John}, and B.~L.~S. {Marlow}, ``Rydberg atom electric field sensors for communications and sensing,'' {\em IEEE Trans. Quantum Eng.}, vol.~2, pp.~1--13, Mar. 2021.

\bibitem{Rydberg_Receiver}
T.~Gong, A.~Chandra, C.~Yuen, Y.~L. Guan, R.~Dumke, C.~M.~S. See, M.~Debbah, and L.~Hanzo, ``Rydberg atomic quantum receivers for classical wireless communication and sensing,'' {\em arXiv preprint arXiv:2409.14501}, 2024.

\bibitem{Atomic-MIMO}
M.~{Cui}, Q.~{Zeng}, and K.~{Huang}, ``Towards atomic {MIMO} receivers,'' {\em arXiv preprint arXiv:2404.04864}, Apr. 2024.

\bibitem{IQ}
M.~Cui, Q.~Zeng, and K.~Huang, ``{MIMO} precoding for {Rydberg} atomic receivers,'' {\em arXiv preprint arXiv:2408.14366}, Oct. 2024.

\bibitem{AtomicTrans}
C.~S. Adams, J.~D. Pritchard, and J.~P. Shaffer, ``Rydberg atom quantum technologies,'' {\em Journal of Physics B: Atomic, Molecular and Optical Physics}, vol.~53, no.~1, p.~012002, 2019.

\bibitem{Rydberg_AoA1}
A.~{Robinson}, N.~{Prajapati}, D.~{Senic}, M.~T. {Simons}, and C.~L. {Holloway}, ``Determining the angle-of-arrival of a radio-frequency source with a {Rydberg} atom-based sensor,'' {\em Appl. Phys. Lett.}, vol.~118, Mar. 2021.

\bibitem{Rydberg_AoA2}
R.~{Mao}, Y.~{Lin}, Y.~{Fu}, Y.~{Ma}, and K.~{Yang}, ``Digital beamforming and receiving array research based on {Rydberg} field probes,'' {\em IEEE Trans. Ant. Propag.}, vol.~72, pp.~2025--2029, Nov. 2023.

\bibitem{Bra_Ket}
T.~Jennewein, U.~Achleitner, G.~Weihs, H.~Weinfurter, and A.~Zeilinger, ``A fast and compact quantum random number generator,'' {\em Rev. Sci. Instrum.}, vol.~71, no.~4, pp.~1675--1680, 2000.

\bibitem{Quantum_Mech}
B.~Zwiebach, {\em Mastering Quantum Mechanics}.
\newblock MIT Press, 2022.

\bibitem{Quantum2}
C.~J. Foot, {\em Atomic physics}, vol.~7.
\newblock Oxford University Press, 2004.

\bibitem{quantum_rotating}
A.~M. Fox, {\em Quantum optics: an introduction}, vol.~15.
\newblock Oxford University Press, USA, 2006.

\bibitem{EIT-AT}
B.~{Liu}, L.~{Zhang}, Z.~{Liu}, Z.~{Deng}, D.~{Ding}, B.~{Shi}, and G.~{Guo}, ``Electric field measurement and application based on {Rydberg} atoms,'' {\em Electromagn. Sci.}, vol.~1, pp.~1--16, Jun. 2023.

\bibitem{Phase_Retrieval}
J.~{Dong}, L.~{Valzania}, A.~{Maillard}, T.~a.~{Pham}, S.~{Gigan}, and M.~{Unser}, ``Phase retrieval: From computational imaging to machine learning: A tutorial,'' {\em IEEE Signal Process. Mag.}, vol.~40, pp.~45--57, Jan. 2023.

\bibitem{Wirtinger}
X.~L. E.~J.~{Candès} and M.~{Soltanolkotabi}, ``Phase retrieval via {Wirtinger} flow: Theory and algorithms,'' {\em IEEE Trans. Inf. Theory}, vol.~61, pp.~1985--2007, Apr. 2015.

\bibitem{MUSIC}
R.~{Schmidt}, ``Multiple emitter location and signal parameter estimation,'' {\em IEEE Trans. Antennas and Propag.}, vol.~34, pp.~276--280, Mar. 1986.

\bibitem{Arc}
N.~{\v{S}}ibali{\'c}, J.~D. Pritchard, C.~S. Adams, and K.~J. Weatherill, ``Arc: An open-source library for calculating properties of alkali {Rydberg} atoms,'' {\em Comput. Phys. Commun.}, vol.~220, pp.~319--331, Nov. 2017.

\bibitem{QSN}
L.~W. Bussey, F.~A. Burton, K.~Bongs, J.~Goldwin, and T.~Whitley, ``Quantum shot noise limit in a {Rydberg} {RF} receiver compared to thermal noise limit in a conventional receiver,'' {\em IEEE Sens. Lett.}, vol.~6, pp.~1--4, Sep. 2022.

\end{thebibliography}

\end{document}